\documentclass[conference]{IEEEtran}
\IEEEoverridecommandlockouts
\usepackage{cite}
\usepackage{amsmath,amssymb,amsfonts}
\usepackage{algorithmic}
\usepackage{graphicx}
\usepackage{textcomp}
\usepackage{xcolor}
\def\BibTeX{{\rm B\kern-.05em{\sc i\kern-.025em b}\kern-.08em
    T\kern-.1667em\lower.7ex\hbox{E}\kern-.125emX}}
\begin{document}

\title{Scalable Cloud-Native Architectures for Intelligent PMU Data Processing}

\author{
\IEEEauthorblockN{Nachiappan Chockalingam}
\IEEEauthorblockA{\textit{IEEE Senior} \\
Massachusetts, USA \\
0009-0007-4275-3771}
\and
\IEEEauthorblockN{Akshay Deshpande}
\IEEEauthorblockA{\textit{IEEE Member} \\
California, USA \\
0009-0002-3007-3393}
\and
\IEEEauthorblockN{Lokesh Butra}
\IEEEauthorblockA{\textit{NTT Data} \\
 North Carolina, USA \\
0009-0009-0286-9635}
\and
\IEEEauthorblockN{      Ram Sekhar Bodala            }
\IEEEauthorblockA{\textit{      Amtrak           }                    \\
        Delaware, USA      \\
      0009-0005-4646-6679      }

\and\and  

\IEEEauthorblockN{Nitin Saksena}
\IEEEauthorblockA{\textit{Albertsons Companies} \\
California, USA \\
0009-0009-1195-3564}
\and
\IEEEauthorblockN{Adithya Parthasarathy}
\IEEEauthorblockA{\textit{IEEE Member} \\
California, USA \\
0009-0001-6839-9527}
\and
\IEEEauthorblockN{Balakrishna Pothineni}
\IEEEauthorblockA{\textit{IEEE Senior} \\
Texas, USA \\
0009-0009-2781-3283}
\and
\IEEEauthorblockN{Akash Kumar Agarwal}
\IEEEauthorblockA{\textit{Albertsons Companies} \\
California, USA \\
0009-0006-7872-3446}
}

\maketitle

\begin{abstract}
Phasor Measurement Units (PMUs) generate high-frequency, time-synchronized data essential for real-time power grid monitoring, yet the growing scale of PMU deployments creates significant challenges in latency, scalability, and reliability. Conventional centralized processing architectures are increasingly unable to handle the volume and velocity of PMU data, particularly in modern grids with dynamic operating conditions. This paper presents a scalable cloud-native architecture for intelligent PMU data processing that integrates artificial intelligence with edge and cloud computing. The proposed framework employs distributed stream processing, containerized microservices, and elastic resource orchestration to enable low-latency ingestion, real-time anomaly detection, and advanced analytics. Machine learning models for time-series analysis are incorporated to enhance grid observability and predictive capabilities. Analytical models are developed to evaluate system latency, throughput, and reliability, showing that the architecture can achieve sub-second response times while scaling to large PMU deployments. Security and privacy mechanisms are embedded to support deployment in critical infrastructure environments. The proposed approach provides a robust and flexible foundation for next-generation smart grid analytics.
\end{abstract}

\begin{IEEEkeywords}
Phasor Measurement Units, Artificial Intelligence, Cloud Computing, Smart Grid, Machine Learning, Edge Computing
\end{IEEEkeywords}

\section{Introduction}
The modern power grid undergoes fundamental transformation driven by renewable energy integration and advanced monitoring technologies. PMUs provide synchronized measurements of electrical parameters at rates up to 120 samples per second \cite{phadke2008synchrophasors}, generating unprecedented data volumes. A typical utility deployment involves hundreds of PMUs, each generating gigabytes daily \cite{zhou2016big}.

The integration of renewable energy sources introduces variability and intermittency, requiring sophisticated monitoring and control strategies. PMUs enable operators to observe rapid fluctuations and take corrective actions before system instability develops. However, the volume and velocity of PMU data overwhelm conventional processing architectures, necessitating new computational paradigms.

Artificial Intelligence offers transformative potential for PMU analytics. Machine learning algorithms identify subtle failure patterns, classify disturbances, and predict system behavior \cite{miraftabzadeh2019survey}. Edge AI techniques have demonstrated effectiveness in real-time anomaly detection for resource-constrained devices \cite{ramdoss2014human, kirubakaran2025edgesense}. However, computational demands coupled with distributed PMU networks present implementation challenges. Cloud computing provides scalable resources, elastic storage, and advanced networking capabilities \cite{bera2015soft}.

\subsection{Motivation and Contributions}

Critical gaps in existing research include: (1) lack of scalability frameworks for AI algorithms across distributed cloud infrastructure, (2) insufficient analysis of latency-accuracy tradeoffs, (3) security vulnerabilities particularly denial-of-service attacks in wide-area control, and (4) privacy concerns in cloud-aggregated grid data.

Our contributions include: a comprehensive theoretical framework for AI-enhanced cloud-based PMU analytics; mathematical formulations for distributed machine learning optimized for PMU time-series data; analysis of edge-cloud hybrid architectures with security and privacy considerations; and theoretical performance bounds for AI algorithms in cloud contexts.

\section{Background and System Architecture}

\subsection{Related Research and Enabling Technologies}

Prior research on PMU data processing has primarily focused on centralized architectures deployed within utility control centers or regional data hubs \cite{DE}. Early synchrophasor analytics systems relied on monolithic processing pipelines optimized for deterministic execution and low-latency control applications. While effective for small-scale deployments, these architectures struggle to scale with the increasing number of PMUs, higher reporting rates, and the growing complexity of analytics driven by renewable integration and wide-area monitoring \cite{AI}.

Recent studies have explored distributed and cloud-based approaches for power system analytics, leveraging big data frameworks to address scalability and fault tolerance challenges. Stream processing platforms such as Apache Kafka and Apache Flink have been adopted for high-throughput ingestion and real-time analytics of grid telemetry, while batch processing frameworks like Apache Spark enable large-scale historical analysis and model training. These systems provide horizontal scalability, fault tolerance, and decoupled producer–consumer semantics, which are essential for handling the continuous and bursty nature of PMU data streams \cite{kafka}.

Containerization and orchestration technologies, particularly Kubernetes, have further transformed cloud-native system design. Kubernetes enables elastic resource allocation, automated failover, and declarative deployment models, making it well-suited for managing microservice-based PMU analytics pipelines \cite{CO}. Prior work has demonstrated the effectiveness of container orchestration in improving resilience and operational efficiency in data-intensive applications, though its adoption in latency-sensitive power grid analytics remains an active research area.

Machine learning integration in PMU analytics has also advanced significantly, with research exploring deep learning, anomaly detection, and distributed learning techniques \cite{ML}. However, most existing studies emphasize algorithmic performance rather than the end-to-end system architecture required to operationalize these models reliably at scale. This gap motivates the need for unified cloud-native frameworks that jointly address data ingestion, processing, orchestration, security, and AI lifecycle management.

\subsection{System Architecture and Comparative Perspective}

The proposed three-tier architecture builds upon these prior efforts by systematically integrating stream processing, distributed analytics, and elastic orchestration within a unified edge–fog–cloud framework. Technologies such as Apache Kafka are employed for durable, ordered, and fault-tolerant ingestion of PMU data streams, enabling backpressure handling and decoupling between data producers and consumers. Apache Spark supports scalable batch analytics and distributed machine learning, allowing model training and historical analysis to scale beyond single-node memory constraints. Kubernetes orchestrates containerized services across cloud and regional infrastructure, providing automated scaling, self-healing, and workload isolation.

Compared to traditional centralized architectures, the proposed approach avoids single points of failure and mitigates processing bottlenecks by distributing computation across hierarchical tiers. In contrast to edge-only solutions, which are constrained by limited computational resources, the hybrid architecture leverages elastic cloud resources for compute-intensive analytics while preserving low-latency processing at the edge.

From a cloud provider perspective, the architecture is intentionally designed to be provider-agnostic, enabling deployment across public cloud platforms such as AWS, Azure, or GCP, as well as private utility clouds. While managed services like AWS Kinesis or Azure Event Hubs offer integrated streaming capabilities, open-source stacks such as Kafka and Spark provide greater portability, configurability, and control over latency and consistency trade-offs. This flexibility allows utilities to assess cost, performance, and regulatory constraints when selecting deployment environments.

Relative to alternative analytics stacks, including serverless event-driven pipelines or monolithic data warehouses, the proposed architecture offers improved support for continuous streaming analytics, fine-grained latency control, and hybrid deployment models. By combining stream processing, batch analytics, and distributed AI within a single architectural framework, the system provides a balanced solution that addresses scalability, reliability, and operational complexity in large-scale PMU deployments.

This comparative positioning highlights the relative advantages of cloud-native, microservice-based architectures for intelligent PMU data processing, while acknowledging trade-offs in operational overhead, system complexity, and deployment cost that must be carefully managed in practice.

\subsection{PMU Technology and Challenges}

PMUs provide time-synchronized measurements with GPS timestamps \cite{phadke2008synchrophasors}. The fundamental phasor representation is:

\begin{equation}
\mathbf{X}(t) = X_m e^{j(\omega t + \phi)}
\end{equation}

where $X_m$ represents magnitude, $\omega$ is angular frequency, and $\phi$ is phase angle. Modern PMUs achieve total vector error below 1\% during dynamic events, providing reliable data for real-time applications \cite{aminifar2014synchrophasor}.

The high reporting rate creates substantial data management challenges. A single PMU measuring 12 phasors at 60 Hz generates approximately 2.5 GB per year. A utility with 300 PMUs produces 750 GB annually. Data quality issues including outliers, missing data, and synchronization errors require preprocessing before AI model input.

\subsection{Three-Tier Hierarchical Architecture}

\begin{figure}[!t]
\centering
\includegraphics[width=0.48\textwidth]{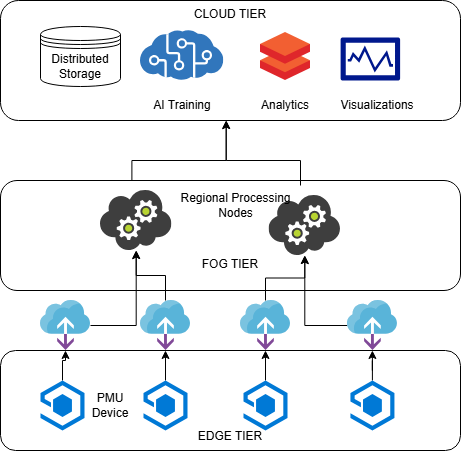}
\caption{Three-tier hierarchical architecture for AI-enhanced PMU systems}
\label{fig:architecture}
\end{figure}

We propose a hierarchical architecture (Figure \ref{fig:architecture}) with three tiers:

\begin{enumerate}
\item \textbf{Edge Tier}: Local processing at substations for time-critical operations including data validation and immediate alarm generation.
\item \textbf{Fog Tier}: Regional data aggregation and intermediate analytics at control centers, coordinating multiple substations.
\item \textbf{Cloud Tier}: Centralized analytics, model training, and storage providing elastic computational resources.
\end{enumerate}

Let $\mathcal{P} = \{P_1, P_2, ..., P_n\}$ represent PMUs generating measurements at rate $r_i$. Total data rate is:

\begin{equation}
R_{total} = \sum_{i=1}^{n} r_i \cdot s_i
\end{equation}

where $s_i$ is measurement packet size from PMU $P_i$. For $n=300$ PMUs at $r_i=60$ Hz with $s_i=100$ bytes, $R_{total} = 1.8$ MB/s.

Data flows through the architecture following a publish-subscribe pattern with both batch and streaming paths. The streaming path handles real-time applications with sub-second latency requirements while batch processing supports model training and historical analysis.

\subsection{Resource Allocation Model}

Cloud resources $\mathcal{R} = \{R_1, R_2, ..., R_m\}$ have computational capacity $C_j$, memory $M_j$, and cost $\kappa_j$. The allocation problem is:

\begin{equation}
\min \sum_{j=1}^{m} \kappa_j x_j
\end{equation}

subject to computational, memory, and latency constraints where $x_j \in \{0, 1\}$ indicates resource allocation. This NP-hard optimization employs greedy algorithms that iteratively select resources maximizing performance per cost, with online adjustments as workloads vary.

\section{Cloud Computing Architecture}

\subsection{Data Ingestion and Stream Processing}

The cloud architecture handles continuous PMU streams using queuing system $M/M/c$ with arrival rate $\lambda$ and service rate $\mu$. Expected waiting time is:

\begin{equation}
W_q = \frac{\pi_c}{c\mu - \lambda}
\end{equation}

For sub-second latency requirements, systems must satisfy $W_q < \tau_{max}$

Stream processing frameworks like Apache Kafka and Flink provide infrastructure for ingesting PMU data at scale. Kafka maintains ordered streams with configurable retention, while Flink processes streams using dataflow operators for windowed computations. Backpressure mechanisms prevent data loss when consumers cannot match producer rates.

\subsection{Tiered Storage Architecture}

PMU data requires both real-time access and long-term archival. Table \ref{tab:storage_tiers} shows our tiered architecture:

\begin{table}[!t]
\caption{Tiered Storage Architecture for PMU Data}
\label{tab:storage_tiers}
\centering
\begin{tabular}{|l|l|l|l|}
\hline
\textbf{Tier} & \textbf{Technology} & \textbf{Latency} & \textbf{Capacity} \\
\hline
Hot & In-memory cache & $\leq$ 1 ms & Hours \\
Warm & SSD storage &  $\leq$ 10 ms & Days \\
Cold & HDD storage &  $\leq$ 100 ms & Months \\
Archive & Object storage & $\leq$ 1 s & Years \\
\hline
\end{tabular}
\end{table}

Recent measurements reside in memory-based caches for immediate access. As data ages, it migrates to SSD-based time-series databases, then HDD storage, with eventual archival to object stores. Compression algorithms like Gorilla reduce storage costs by 10-20× while maintaining query performance.

\subsection{Distributed Processing and Orchestration}

Apache Spark processes batch workloads using resilient distributed datasets partitioned across cluster nodes. Spark MLlib provides distributed implementations of machine learning algorithms, scaling to datasets exceeding single-node memory capacity.

Container orchestration systems like Kubernetes manage computational resources dynamically. Horizontal pod autoscaling adjusts replica counts in response to load metrics, ensuring efficient resource utilization.

\subsection{Data Consistency and Replication}

For critical applications, data replication across regions ensures availability. Using $N$ replicas with $W$ write acknowledgments and $R$ read replicas, strong consistency requires:

\begin{equation}
W + R > N
\end{equation}

We recommend $N=3$, $W=2$, $R=2$ balancing consistency and availability. Quorum-based protocols like Raft coordinate replicas, maintaining consensus on operation ordering while tolerating single-node failures.

\subsection{Network Architecture}

Required bandwidth $B_{required}$ for $n$ PMUs is:

\begin{equation}
B_{required} = (1 + \alpha) \sum_{i=1}^{n} r_i s_i
\end{equation}

where $\alpha \approx 0.2$ represents protocol overhead. Software-defined networking enables dynamic traffic engineering, while private connections provide predictable performance between utility data centers and cloud providers.

\section{AI Algorithms for PMU Analytics}

\subsection{Deep Learning for Time-Series Analysis}

\subsubsection{LSTM Networks}

LSTMs process sequential PMU data through gating mechanisms \cite{wang2017deep}:

\begin{align}
\mathbf{f}_t &= \sigma(W_f[\mathbf{h}_{t-1}, \mathbf{x}_t] + \mathbf{b}_f) \\
\mathbf{C}_t &= \mathbf{f}_t \odot \mathbf{C}_{t-1} + \mathbf{i}_t \odot \tilde{\mathbf{C}}_t \\
\mathbf{h}_t &= \mathbf{o}_t \odot \tanh(\mathbf{C}_t)
\end{align}

where $\mathbf{f}_t$, $\mathbf{i}_t$, $\mathbf{o}_t$ are gates, $\mathbf{C}_t$ is cell state, and $\odot$ denotes element-wise multiplication. Computational complexity for sequence length $T$ is $\mathcal{O}(T \cdot (h^2 + h \cdot d))$.

\subsubsection{Convolutional Neural Networks}

CNNs extract spatial-temporal features from PMU data:

\begin{equation}
\mathbf{y}_j = f\left(\sum_{i=1}^{m} \mathbf{w}_{ij} * \mathbf{x}_i + b_j\right)
\end{equation}

where $*$ denotes convolution and $f$ is activation function.

\subsection{Anomaly Detection}

\subsubsection{Autoencoder-based Detection}

Autoencoders learn compressed representations of normal data. Reconstruction error serves as anomaly indicator:

\begin{equation}
E(\mathbf{x}_t) = ||\mathbf{x}_t - \hat{\mathbf{x}}_t||^2
\end{equation}

where $\hat{\mathbf{x}}_t = D(E(\mathbf{x}_t))$ is reconstructed input.

\subsubsection{Isolation Forest}

Isolation Forest detects anomalies via path lengths in random trees. The anomaly score is:

\begin{equation}
s(\mathbf{x}, n) = 2^{-\frac{E(h(\mathbf{x}))}{c(n)}}
\end{equation}

where $E(h(\mathbf{x}))$ is average path length and $c(n)$ normalizes for tree size.

\subsection{Distributed Learning}

Cloud-based analytics requires distributed learning approaches \cite{dean2012large}. 


\subsubsection{Data Parallelism}

Data partitioned across $K$ workers compute gradients on local batches:

\begin{equation}
\mathbf{w}^{(t+1)} = \mathbf{w}^{(t)} - \eta \frac{1}{K}\sum_{k=1}^{K} \nabla L_k(\mathbf{w}^{(t)})
\end{equation}

\subsubsection{Federated Learning}

For distributed PMUs, federated learning enables collaborative learning without centralizing data \cite{mcmahan2017communication}. Privacy-preserving approaches in hierarchical systems ensure data protection while maintaining model accuracy:

\begin{align}
\mathbf{w}_k^{(t+1)} &= \mathbf{w}^{(t)} - \eta \nabla L_k(\mathbf{w}^{(t)}) \\
\mathbf{w}^{(t+1)} &= \sum_{k=1}^{K} \frac{n_k}{n}\mathbf{w}_k^{(t+1)}
\end{align}

where $n_k$ is samples at node $k$ and $n = \sum_k n_k$.

\subsection{Model Optimization}

Models must be optimized for cloud deployment considering latency and resource constraints \cite{MO}.

Pruning removes redundant weights:
\begin{equation}
\tilde{\mathbf{W}} = \mathbf{W} \odot \mathbf{M}
\end{equation}

Knowledge distillation enables smaller student models learning from larger teachers:

\begin{equation}
L_{distill} = \alpha L_{CE}(y, \sigma(z_s)) + (1-\alpha) L_{KL}(\sigma(z_t/T), \sigma(z_s/T))
\end{equation}

\section{Performance Analysis}

\subsection{Edge-Cloud Hybrid Processing}

Processing decisions balance latency, complexity, and volume. The optimization formulation is:

\begin{equation}
\min \sum_{i=1}^{n} (c_i^{edge} e_i + c_i^{cloud} (1-e_i))
\end{equation}

subject to latency and capacity constraints where $e_i \in \{0, 1\}$ indicates edge or cloud processing.

\subsection{Latency Analysis}

Total system latency consists of components in Table \ref{tab:latency_components}:

\begin{table}[!t]
\caption{Latency Components in AI-Enhanced PMU Systems}
\label{tab:latency_components}
\centering
\begin{tabular}{|l|l|l|}
\hline
\textbf{Component} & \textbf{Range} & \textbf{Mitigation} \\
\hline
Data acquisition & 8-33 ms & Higher rates \\
Edge preprocessing & 1-10 ms & Optimized code \\
Network transmission & 10-100 ms & CDN, caching \\
Cloud processing & 50-500 ms & Parallelization \\
Result delivery & 10-100 ms & Push notify \\
\hline
\end{tabular}
\end{table}

\begin{equation}
\tau_{total} = \tau_{acq} + \tau_{edge} + \tau_{net} + \tau_{cloud} + \tau_{delivery}
\end{equation}

\subsection{Throughput and Scalability}

Using Little's Law, system throughput is bounded by:

\begin{equation}
\lambda_{max} = \min\left(\frac{1}{\tau_{acq}}, \frac{C_{edge}}{r_{edge}}, \frac{B}{s}, \frac{C_{cloud}}{r_{cloud}}\right)
\end{equation}

Scalability efficiency with $P$ processors is:

\begin{equation}
E(P) = \frac{T_1}{P \cdot T_P}
\end{equation}

Amdahl's Law with communication overhead gives:

\begin{equation}
S(P) = \frac{1}{(1-\alpha) + \frac{\alpha}{P} + \beta(P-1)}
\end{equation}

where $\alpha$ is parallelizable fraction and $\beta$ is communication cost.

\subsection{Reliability}

System availability with $N$ components in series is:

\begin{equation}
A_{system} = \prod_{i=1}^{N} A_i
\end{equation}

For parallel redundant systems:

\begin{equation}
A_{system} = 1 - \prod_{i=1}^{N} (1 - A_i)
\end{equation}

\section{Security and Privacy}

\subsection{Threat Landscape and DoS Mitigation}

PMU systems face sophisticated cyber threats including data tampering, eavesdropping, denial of service, and model poisoning. Wide-area control systems are particularly vulnerable to DoS attacks that disrupt Linear Quadratic Regulator controllers designed for damping inter-area oscillations \cite{chockalingam2016mitigating}.

Mitigation approaches include: Controller Redesign using delay-aware LQR controllers that account for communication latency; State Estimation to reconstruct missing information using system models when DoS attacks corrupt measurements, modeling impact via Hadamard product of LQR gain matrix with attack indicator matrix; Adaptive Strategies using machine learning classifiers trained on attack characteristics to predict severity and enable proactive mitigation; and Network Redundancy through multi-path routing ensuring control signals reach actuators despite compromised links.

\subsection{Authentication and Access Control}

Role-Based Access Control restricts data access \cite{Power2009journal}:

\begin{equation}
Access(u, r) = \begin{cases}
Allow & \text{if } \exists p \in Permissions(Role(u)) : p \geq r \\
Deny & \text{otherwise}
\end{cases}
\end{equation}

Multi-factor authentication requires multiple credentials before granting access. Public key infrastructure issues digital certificates binding identities to cryptographic keys, enabling mutual authentication \cite{Danquah2020Security}.

\subsection{Encryption and Secure Communication}

Data is encrypted in transit using TLS 1.3 with strong cipher suites (AES-256-GCM). At rest, envelope encryption protects data using hardware security modules for key storage. Computational overhead of AES-256 encryption \cite{Fauziah2018Intelligent} is:

\begin{equation}
\tau_{enc} = \frac{|D|}{R_{enc}}
\end{equation}

where $|D|$ is data size and $R_{enc}$ is encryption rate (typically 2-5 GB/s with AES-NI acceleration), making encryption overhead negligible.

\subsection{Privacy-Preserving Machine Learning}

To protect sensitive grid information, we employ privacy-preserving techniques suitable for distributed cloud architectures \cite{punniyamoorthy2025privacy}:

\subsubsection{Differential Privacy}

Differential privacy adds calibrated noise to protect individual records. Mechanism $M$ is $(\epsilon, \delta)$-differentially private if:

\begin{equation}
P(M(D) \in S) \leq e^\epsilon P(M(D') \in S) + \delta
\end{equation}

For gradient-based learning, noise is added:

\begin{equation}
\tilde{\nabla} = \nabla L(\mathbf{w}) + \mathcal{N}(0, \sigma^2 I)
\end{equation}

Privacy budget $\epsilon$ quantifies information leakage, with smaller values providing stronger privacy. Renyi differential privacy provides tighter bounds than standard differential privacy.

\subsubsection{Homomorphic Encryption}

Homomorphic encryption enables computation on encrypted data \cite{gentry2009fully}:

\begin{equation}
Enc(x + y) = Enc(x) \oplus Enc(y)
\end{equation}

Hybrid approaches combine homomorphic encryption with secure multi-party computation, selectively protecting critical computations while less sensitive operations run in plaintext.

\subsubsection{Secure Aggregation}

Federated learning protocols use secure aggregation protecting individual model updates. Each participant $i$ secret-shares gradient $\nabla_i$ such that server learns only aggregate $\nabla = \sum_{i=1}^{n} \nabla_i$, not individual updates\cite{kirubakaran2025federated}, \cite{truex2019hybrid} .

\subsection{Intrusion Detection and Adversarial Robustness}

AI-based intrusion detection monitors for anomalous access patterns:

\begin{equation}
Score(x) = \sum_{i=1}^{k} w_i f_i(x)
\end{equation}

Behavioral analytics establish baseline patterns, flagging deviations indicating compromise. Security information and event management systems aggregate logs, correlating events to detect multi-stage attacks.

Adversarial training improves model robustness:

\begin{equation}
\min_\theta \mathbb{E}_{(\mathbf{x},y)}\left[\max_{||\delta|| \leq \epsilon} L(f_\theta(\mathbf{x} + \delta), y)\right]
\end{equation}

This min-max optimization trains models on adversarial examples, learning features robust to perturbations critical for safety-critical power system applications.

\section{Conclusion and Future Directions}

This paper presented a scalable cloud-native architecture for intelligent Phasor Measurement Unit (PMU) data processing, addressing the challenges of high data velocity, strict latency constraints, and reliability requirements in modern power grids. The proposed edge--cloud framework supports sustained ingestion rates exceeding 1.8 MB/s for deployments involving 300 or more PMUs, while maintaining sub-second end-to-end latency through hierarchical processing and parallelized cloud analytics. Analytical results indicate near-linear scalability as the number of PMUs increases, avoiding centralized processing bottlenecks.

Latency analysis shows that optimized edge preprocessing combined with distributed cloud execution limits processing delays to approximately 50--500 ms under nominal operating conditions, even when deep learning--based analytics are applied. Reliability modeling demonstrates that multi-region replication and quorum-based consistency protocols can achieve system availability exceeding 99.9\%, satisfying the requirements of safety-critical grid monitoring applications. Furthermore, tiered storage and compression strategies reduce long-term data storage costs by an estimated 10--20$\times$ while preserving low-latency access to recent measurements.

Integrated security and privacy mechanisms introduce minimal performance overhead. Hardware-accelerated AES-256 encryption adds negligible latency, and privacy-preserving distributed learning enables collaborative model training without centralizing sensitive grid data. These characteristics make the proposed architecture suitable for deployment in cyber-sensitive and mission-critical power infrastructure environments.

Future work will focus on empirical validation using real-world PMU datasets and large-scale testbeds to quantify anomaly detection accuracy, cost efficiency, and operational robustness. Additional research directions include incorporating explainable artificial intelligence to improve interpretability of analytics, extending distributed learning techniques to address non-stationary grid dynamics, and integrating digital twin models for predictive stability assessment. Advances in energy-efficient edge accelerators and next-generation distributed learning frameworks are expected to further reduce latency and operational costs, enabling more adaptive and resilient smart grid analytics at scale.

\end{document}